\documentclass[letterpaper]{article}
\usepackage[sort&compress]{natbib}
\usepackage{alifeconf}

\usepackage{calc}

\usepackage[shortcuts]{extdash}

\let\oldtitle\title
\renewcommand\title[1]{%
    \hypersetup{pdftitle={#1}}%
    \oldtitle{#1}%
    \def\thetitle{#1}%
    \pdfbookmark[0]{#1}{title}}

\newcommand\andnext{\unskip, }%
\newcommand\authoraffils[1]{\mbox{}\\\mbox{}\\#1}
\newcommand\corremail[1]{\mbox{}\\\emaillink{#1}}
\let\oldauthor\author
\renewcommand\author[1]{%
    \begingroup
        \providecommand\and{}%
        \renewcommand\and{\unskip, }%
        \providecommand\andnext{}%
        \renewcommand\andnext{\unskip, }%
        \providecommand\textsuperscript[1]{}%
        \renewcommand\textsuperscript[1]{}%
        \providecommand\authoraffils[1]{}%
        \renewcommand\authoraffils[1]{}%
        \providecommand\corremail[1]{}%
        \renewcommand\corremail[1]{}%
        \hypersetup{pdfauthor={#1}}%
        \xdef\theauthor{#1}%
    \endgroup
    \oldauthor{#1}%
    }

\usepackage{graphicx}

\usepackage{booktabs}

\usepackage{xspace}

\makeatletter
\def\citet{\@ifstar{\citetstar}{\citetnostar}}
\def\Citet{\@ifstar{\Citetstar}{\Citetnostar}}
\def\citetnostar{\@ifnextchar[{\squarecitet}{\simplecitet}}
\def\squarecitet[#1]{\@ifnextchar[{\twocitet[#1]}{\onecitet[#1]}}
\def\Citetnostar{\@ifnextchar[{\squareCitet}{\simpleCitet}}
\def\squareCitet[#1]{\@ifnextchar[{\twoCitet[#1]}{\oneCitet[#1]}}
\def\citetstar{\@ifnextchar[{\squarecitetstar}{\simplecitetstar}}
\def\squarecitetstar[#1]{\@ifnextchar[{\twocitetstar[#1]}{\onecitetstar[#1]}}
\def\Citetstar{\@ifnextchar[{\squareCitetstar}{\simpleCitetstar}}
\def\squareCitetstar[#1]{\@ifnextchar[{\twoCitetstar[#1]}{\oneCitetstar[#1]}}
\makeatother
\def\simplecitet#1{\citeauthor{#1}~\citeyearpar{#1}}
\def\onecitet[#1]#2{\citeauthor{#2}~\citeyearpar[#1]{#2}}
\def\twocitet[#1][#2]#3{\citeauthor{#3}~\citeyearpar[#1][#2]{#3}}
\def\simplecitetstar#1{\citeauthor*{#1}~\citeyearpar{#1}}
\def\onecitetstar[#1]#2{\citeauthor*{#2}~\citeyearpar[#1]{#2}}
\def\twocitetstar[#1][#2]#3{\citeauthor*{#3}~\citeyearpar[#1][#2]{#3}}
\def\simpleCitet#1{\Citeauthor{#1}~\citeyearpar{#1}}
\def\oneCitet[#1]#2{\Citeauthor{#2}~\citeyearpar[#1]{#2}}
\def\twoCitet[#1][#2]#3{\Citeauthor{#3}~\citeyearpar[#1][#2]{#3}}
\def\simpleCitetstar#1{\Citeauthor*{#1}~\citeyearpar{#1}}
\def\oneCitetstar[#1]#2{\Citeauthor*{#2}~\citeyearpar[#1]{#2}}
\def\twoCitetstar[#1][#2]#3{\Citeauthor*{#3}~\citeyearpar[#1][#2]{#3}}

\usepackage{amsfonts,amssymb}
\usepackage{mathrsfs}                   

\usepackage{varioref}
\labelformat{equation}{\textup{(#1)}}

\usepackage[%
        pagebackref=false,
        bookmarks=true,bookmarksopen=true,
        bookmarksnumbered=true,
        breaklinks=true,
        colorlinks=true,
        anchorcolor=,citecolor=,
        urlcolor=,linkcolor=,filecolor=,
        menucolor=,
        pdfpagelabels=true,hypertexnames=false,
        plainpages=false,
        ]{hyperref}

\usepackage{url}
\usepackage{doi}

\usepackage{subfigure}

\makeatletter
\@ifpackageloaded{hyperref}{%
\@ifpackageloaded{amsmath}{%
\newcommand{\AMShreffix}[1]{%
        \expandafter\let\csname AMShreffix#1\expandafter\endcsname%
                \csname #1\endcsname%
        \expandafter\renewcommand\csname #1\endcsname{%
                \@hyper@itemfalse\csname AMShreffix#1\endcsname}}
\AtBeginDocument{%
\AMShreffix{equation}
\AMShreffix{align}
\AMShreffix{alignat}
\AMShreffix{flalign}
\AMShreffix{gather}
\AMShreffix{multline}
}}{}}{}
\makeatother

\let\orgautoref\autoref

\def\Enospace~{}

\renewcommand{\autoref}
        {\def\equationautorefname{Eq.}%
         \def\figureautorefname{Fig.}%
         \def\sectionautorefname{Section}%
         \def\appendixautorefname{Appendix}%
         \def\Itemautorefname{item}%
         \def\tableautorefname{Table}%
         \orgautoref}

\newcommand\insilico{\emph{in silico}\xspace}
\newcommand\ie{i.e.\xspace}

\title{Self-referencing cellular automata: A model of the evolution of
    information control in biological systems}
\author{Theodore P.~Pavlic\textsuperscript{1}%
            \andnext
            Alyssa M.~Adams\textsuperscript{1}%
            \andnext
            Paul C.~W.~Davies\textsuperscript{1}%
            \and
            Sara Imari Walker\textsuperscript{1}%
    \authoraffils{\textsuperscript{1}Arizona State University, Tempe, AZ  85287}%
    \corremail{tpavlic@asu.edu}}

\begin{document}
\maketitle

\begin{abstract}
    Cellular automata have been useful artificial models for
    exploring how relatively simple rules combined with spatial memory
    can give rise to complex emergent patterns. Moreover, studying the
    dynamics of how rules emerge under artificial selection for function
    has recently become a powerful tool for understanding how evolution
    can innovate within its genetic rule space. However, conventional
    cellular automata lack the kind of state feedback that is surely
    present in natural evolving systems. Each new generation of a
    population leaves an indelible mark on its environment and thus
    affects the selective pressures that shape future generations of
    that population. To model this phenomenon, we have augmented
    traditional cellular automata with state\-/dependent feedback.
    Rather than generating automata executions from an initial condition
    and a static rule, we introduce mappings which generate iteration
    rules from the cellular automaton itself. We show that these new
    automata contain disconnected regions which locally act like
    conventional automata, thus encapsulating multiple functions into
    one structure. Consequently, we have provided a new model for
    processes like cell differentiation. Finally, by studying the size
    of these regions, we provide additional evidence that the dynamics
    of self\-/reference may be critical to understanding the evolution
    of natural language. In particular, the rules of elementary cellular
    automata appear to be distributed in the same way as words in the
    corpus of a natural language.
\end{abstract}

\section{Introduction}

Cellular automata~(CA) are model complex systems that combine spatial
memory with relatively simple update rules to produce rich dynamic
patterns~\citep{Wolfram02}. In this regard, they can be viewed as models
for life. ``Rules'' in DNA encode policies that iteratively react to the
environment by modifying it. Thus, over evolutionary time scales,
natural selection can explore the nucleic\-/acid rule space and amplify
those rules which provide useful functions. With this narrative in mind,
researchers have used \insilico{} artificial selection to explore the CA
rule space and amplify rules with certain computational
abilities~\citep{DMC94, MCH94, BB05, Hordijk13}. Moreover, there has
been much interest in understanding the demographic dynamics of these CA
rule populations over their evolutionary history~\citep{MCH94,
Hordijk13}. That is, artificial selection of these evolving CA's has
itself become a model for the innovation intrinsic to natural selection.

One major difference between evolving cellular automata~(EvCA) and
evolving natural organisms is the lack of feedback in the fitness
channel of the former. In EvCA, each new generation faces the same
selective pressures as prior generations. However, with new generations
of natural organisms, there is feedback between the current demographics
and the selective pressures shaping future demographics. \Citet{GW11}
point out that these self\-/referential dynamics are a unique
characteristic of life~-- making life distinctly different from any
other physical system. Two of us~(SIW and~PCWD) have proposed that
self\-/referential dynamics are one of the hallmarks of life, emerging
with its origin~\citep{WD13}. Where EvCA's will only innovate in the
presence of external ``abiotic'' pressures, natural organisms put
pressure on themselves to re\-/organize even without an external fitness
driver. Similarly, at everyday and ontogenetic time scales, conventional
CA's do not easily embed the regulatory mechanisms that permeate
throughout life. A single genome gives rise to a wide variety of
differentiated cell phenotypes which, locally, appear to follow a
consistent set of operational rules but globally appear to have no
shared program. By modeling feedback explicitly, these phenomena can be
explained using gene regulatory network~(GRN)
frameworks~\citep{SchlittBrazma07, Davidson10}, where the expression
level of one gene promotes or inhibits the expression level of another
and thus ``latches'' cells into different types. Thus, by making
feedback between state and dynamics explicit in CA's, it may be possible
to enrich their ability to model how organisms emerge, evolve, develop,
and react to both their external environment and their internal state.

Here, we introduce self\-/referencing cellular\-/automaton framework we
call PICARD where \emph{\textbf{P}ICARD \textbf{I}mplements \textbf{CA}
\textbf{R}ules \textbf{D}ifferently}~(PICARD). Like a traditional
one\-/dimensional CA, PICARD executions move from one iteration to
another by some rule. However, whereas traditional CA's require the rule
to be static and externally specified, PICARD infers the iteration rule
from the current state of the CA itself. As we will show, executions
from multiple static CA's can be embedded within a single PICARD~-- a
PICARD can be identical to one static CA from certain initial conditions
and another static CA from other initial conditions. Thus, a PICARD can
combine the computational abilities of different CA's within one entity.
Moreover, whereas CA's differ by their rule, PICARD's differ by their
state-to-rule mapping. Because there are many more state-to-rule
mappings than there are CA rules, the PICARD parameter space is
potentially much richer for later evolutionary investigations.

To some extent, PICARD is a simple attempt to add dynamical feedback
that is missing in traditional evolutionary cellular automata. However,
because iterations are generated by rules that are encoded in previous
iterations, PICARD feedback is the kind of self\-/reference that is
thought to be a characteristic feature of life~\citep{Hofstadter79,
KK00a, KK00b, GW11, WD13}. Our CA approach shares many similarities with
self\-/referencing functional\-/dynamics developed by \citeauthor{KK00a}
to model the evolution of rules~\citep{KK00a, KK00b}. Attempting to
avoid the biochemical complexities of the evolution of nucleic acids,
they turn their focus on the evolution of natural language. Furthermore,
they draw connections between attractors in their
coupled\-/logistic\-/map landscape and words that accumulate in
language. Although their framework is very different than the automata
we study here, we too have results that appear to be strongly connected
to the evolution of natural language. Thus, augmenting cellular automata
with self\-/reference widens our ability to model to evolution of
language in unanticipated ways.

\section{Cellular Automata with PICARD Mappings}

\begin{figure}\centering
    \subfigure[Single iteration of rule 182]{\label{fig:traditional_CA_summary}\includegraphics[trim={0pt 10pt 0pt 5pt}, clip, width=\columnwidth]{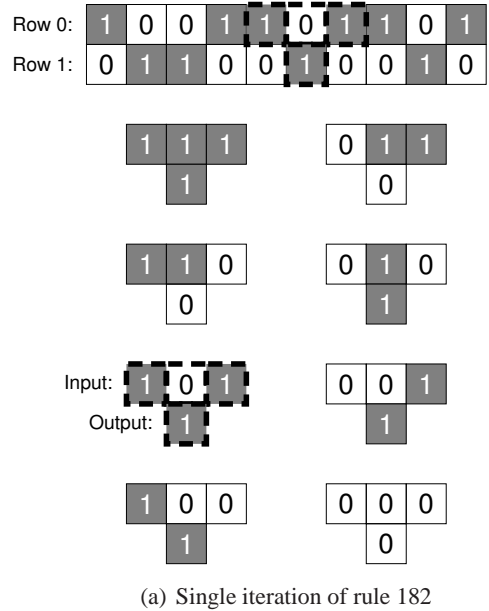}}\\
    \subfigure[Execution of rule 182]{\label{fig:traditional_CA_execution}\includegraphics[trim={0pt 20pt 0pt 0pt}, clip, width=\columnwidth]{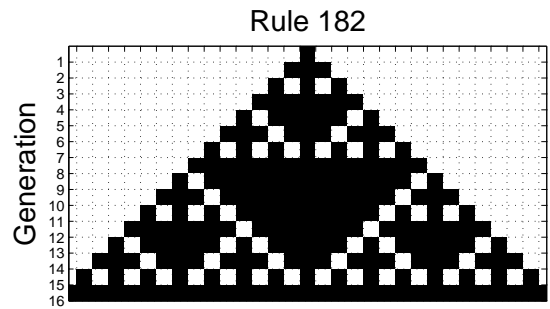}}
    \caption
        [Traditional cellular automaton.]
        {Traditional cellular automaton. In
        \subref{fig:traditional_CA_summary}, a single iteration of a
        traditional CA is shown in the top two rows, where row~0 is
        taken as an initial condition on which a static CA rule operates
        to produce row~1. The static CA rule is summarized in the eight
        groups of four cells in the bottom of the figure. In each group,
        the top three cells match adjacent cells in row~0, and the
        bottom one cell represents the cell generated beneath the row-0
        cells in row~1. One such grouping in the two rows is outlined in
        a broken line along with the corresponding group from the
        elementary CA rule below it. This CA rule can be equivalently
        summarized as an eight-bit binary string 0b10110110, a
        two\-/digit hexadecimal 0xB6, or a decimal 182. In
        \subref{fig:traditional_CA_execution}, an execution of rule~182
        is shown that resembles a Sierpinski triangle.}
    \label{fig:traditional_CA}
\end{figure}

Although PICARD implements CA rules differently, once each rule is
defined, a PICARD iteration is identical to an iteration of a
conventional elementary CA. As shown in
\autoref{fig:traditional_CA_summary}, a traditional elementary CA
generates each row based on the pattern in the preceding row.
The iteration rule is a lookup table that maps each triplet of bits in
the preceding row to a single bit in the following row. Thus, with the
right initial conditions, some rules can produce intricate patterns over
many generations, as shown in \autoref{fig:traditional_CA_execution}.

Where PICARD differs from a traditional CA is that no static rule is
specified. Instead, a map is provided from each row to the rule that
will operate on it. We are primarily interested in CA's with more than
eight cells; consequently, this mapping is a coarse graining of the
system~-- some rules will necessarily correspond to multiple different
row patterns. With this multiple realizability of rules in mind, each
PICARD row and rule can be viewed as a
\emph{microstate}\--\emph{macrostate} pair. Consequently, in the
following, we will use the terms \emph{rule} and \emph{macrostate}
interchangeably.

\begin{figure}\centering
    \subfigure[Sum of ones]{\label{fig:example_mapping_sum}\includegraphics[width=\columnwidth]{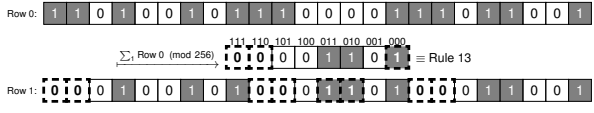}}\\
    \subfigure[Density of ones]{\label{fig:example_mapping_density}\includegraphics[width=\columnwidth]{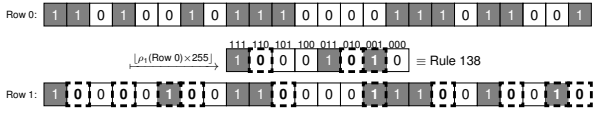}}
    \caption
        [Example aggregation PICARD mappings.]
        {Example aggregation PICARD mappings. The CA rule induced by
        each mapping is shown between rows~0 and~1, where the triplet
        above each position represents the pattern of bits in row~0 that
        will result in the boxed bit in row~1. Positions in the rule
        that result in a changed bit from row~0 to row~1 have been
        highlighted with broken lines; the resulting toggled bits in
        row~1 have also been highlighted. Each mapping represents some
        aggregate property of the first row.
        In \subref{fig:example_mapping_sum}, the rule represents
        the total number of 1's in row~0, modulo 256. Because there are
        only 24 cells in the row, this PICARD mapping can only induce 25
        different elementary CA rules. In the case of a row with more
        than 255 cells, every elementary rule is possible. In
        \subref{fig:example_mapping_density}, the rule
        represents the density of 1's in row~0 scaled by 255.}
    \label{fig:example_aggregation_mappings}
\end{figure}

For example, \autoref{fig:example_aggregation_mappings} shows two
arbitrary examples of microstate\-/to\-/rule mappings.
Each of these two mappings extract an aggregate property of the
corresponding row. In \autoref{fig:example_mapping_sum}, the focal
macrostate is the number of 1's in the row. In
\autoref{fig:example_mapping_density}, the focal macrostate is the
density of 1's in the row. For the analogous case of interacting
particles, macroscopic aggregates like temperature and pressure are not
typically viewed as having causal influence on the microscopic states of
the system. That is, macroscopic states and microscopic states exist in
two different levels of description. However, if a collection of
particles is put into contact with a heat bath characterized
by only its temperature, the macroscopic properties of the gas appear to
drive the evolution of the microscopic system. Likewise, these PICARD
aggregation mappings are simultaneously descriptive coarse grainings as
well as prescriptive rules governing the dynamics of the cells.
\Citet{GW11} argue that self\-/reference appears in biological systems
but apparently not in physical systems because of similar reasoning~--
biological phenomena are emergent and may require self\-/reference for
analysis without appealing to underlying interacting physical
microstates. Self\-/reference therefore may be intimately related to
framework suggested to characterize emergence, such as top-down
causation~\citep{Davies12}, and may play an important role in the
emergence of new organizational levels through major evolutionary
transitions~\citep{WCD12}.

\begin{figure}\centering
    \subfigure[Block ones majority]{\label{fig:example_mapping_majority}\includegraphics[width=\columnwidth]{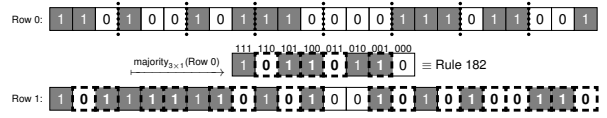}}\\
    \subfigure[Block ones odd parity]{\label{fig:example_mapping_odd}\includegraphics[width=\columnwidth]{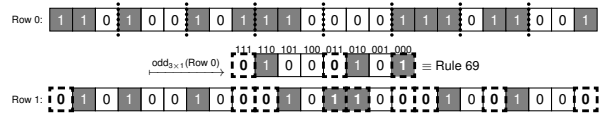}}\\
    \subfigure[Block transition odd parity]{\label{fig:example_mapping_transition}\includegraphics[width=\columnwidth]{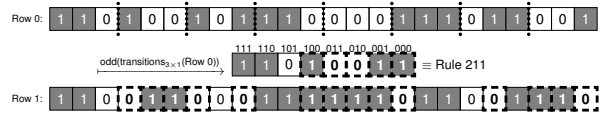}}
    \caption
        [Example block PICARD mappings.]
        {Example block PICARD mappings. Rules are shown just as in
         \protect\autoref{fig:example_aggregation_mappings}. However,
         the 24 bits in row~0 have been broken into 8 3-bit clusters
         shown with separating fences. The eight bits of the rule are
         then induced from a simple 3-bit predicate from each group to
         the corresponding rule bit.
         In \subref{fig:example_mapping_majority}, each bit in the rule
         is a 1 if and only if the corresponding 3-bit cluster in row~0
         has a majority of 1's.
         In \subref{fig:example_mapping_odd}, the rule bit is set if
         there are an odd number of 1's in the cluster.
         In \subref{fig:example_mapping_transition}, the rule bit is set
         if there are an odd number of zero\-/to\-/one or
         one\-/to\-/zero transitions in the 3-bit cluster (see
         \protect\autoref{tab:transition_lookup_table}).}
    \label{fig:example_block_mappings}
\end{figure}

Aggregation mappings may have intuitive connections to statistical
mechanics, but PICARD mappings may be built in entirely different ways.
The block mappings in \autoref{fig:example_block_mappings} are formed by
clustering bits of the CA microstate into eight groups and matching each
group to a predicate function that determines the bit in the rule which
has the same relative position as the 3-bit group.
\begin{table}\centering
    \begin{tabular}{ccc}
        \toprule
        Microstate Block & Number of Transitions & Rule Bit\\
        \midrule
        000 & 0 & 0 \\
        001 & 1 & 1 \\
        010 & 2 & 0 \\
        011 & 1 & 1 \\
        100 & 1 & 1 \\
        101 & 2 & 0 \\
        110 & 1 & 1 \\
        111 & 0 & 0 \\
        \bottomrule
    \end{tabular}
    \caption
        [Lookup table for transition odd parity.]
        {Lookup table for transition odd parity. Reading from left to
        right, the number of zero\-/to\-/one and one\-/to\-/zero
        transitions is counted. The odd parity of this count is
        equivalent to the exclusive or of the outer two bits in the
        microstate block.}
    \label{tab:transition_lookup_table}
\end{table}
\begin{figure}\centering
    \includegraphics[width=\columnwidth]{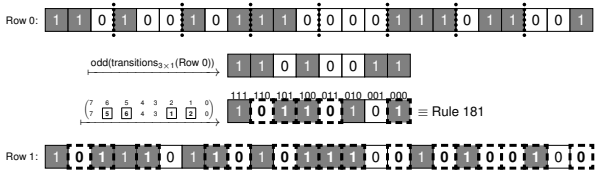}
    \caption
        [Block transition mapping with permutation.]
        {Block transition mapping with permutation. This PICARD mapping
         is equivalent to the mapping from
         \protect\autoref{fig:example_mapping_transition} composed with
         a permutation that switches middle two bits of each nybble.}
    \label{fig:example_permuted_transition}
\end{figure}
For example, in \autoref{fig:example_mapping_majority}, the 24 bits of
the microstate are broken into eight mutually exclusive groups of three
bits. Each 3-bit group is replaced by a single 1 if it has a majority
of~1's and a~0 otherwise. In \autoref{fig:example_mapping_odd}, groups
are mapped into~1 if they have an odd number of~1's. Alternatively, in
\autoref{fig:example_mapping_transition}, the actual binary identity of
the bits in each group is ignored and instead the number of transitions
is used (see \autoref{tab:transition_lookup_table}). That is, from left
to right within each 3-bit group, a cell will transition from one level
to another~0, 1, or~2 times. The corresponding rule bit will be a zero
unless the 3-bit group only contains a single transition. This predicate
is equivalent to the exclusive or~(XOR) of the outer two bits in the
microstate group. As these examples show, PICARD executions starting
from the same microstate can be very different based on the chosen
mapping from microstate to rule.

While the predicates used within block mappings may be relatively
simple, the selection of microstate and macrostate bits for each block
mapping adds an additional layer of complexity. For example, a single
PICARD block mapping may use different predicates for each microstate
block. Additionally, microstate blocks may not be mutually exclusive~--
they may overlap or leave some microstate bits uncovered. Even when a a
single predicate is used over a set of mutually exclusive, equally sized
microstate groups that cover all microstate bits, the ordering of the
microstate groups need not match the ordering of the corresponding rule
bits. In \autoref{fig:example_permuted_transition}, a mapping of this
sort is shown.
That is, the mapping from \autoref{fig:example_mapping_transition} has
been composed with a permutation. For the remainder of this paper, we
will use this mapping to demonstrate the richness of an individual
PICARD.

\section{A PICARD Case Study}

\begin{figure}\centering
    \subfigure[Execution leading to mutual oscillation]{%
        \label{fig:mutual_oscillation}%
        \centerline{%
        \includegraphics[trim={0pt 30pt 0pt -10pt}, clip, height=0.25\textheight]{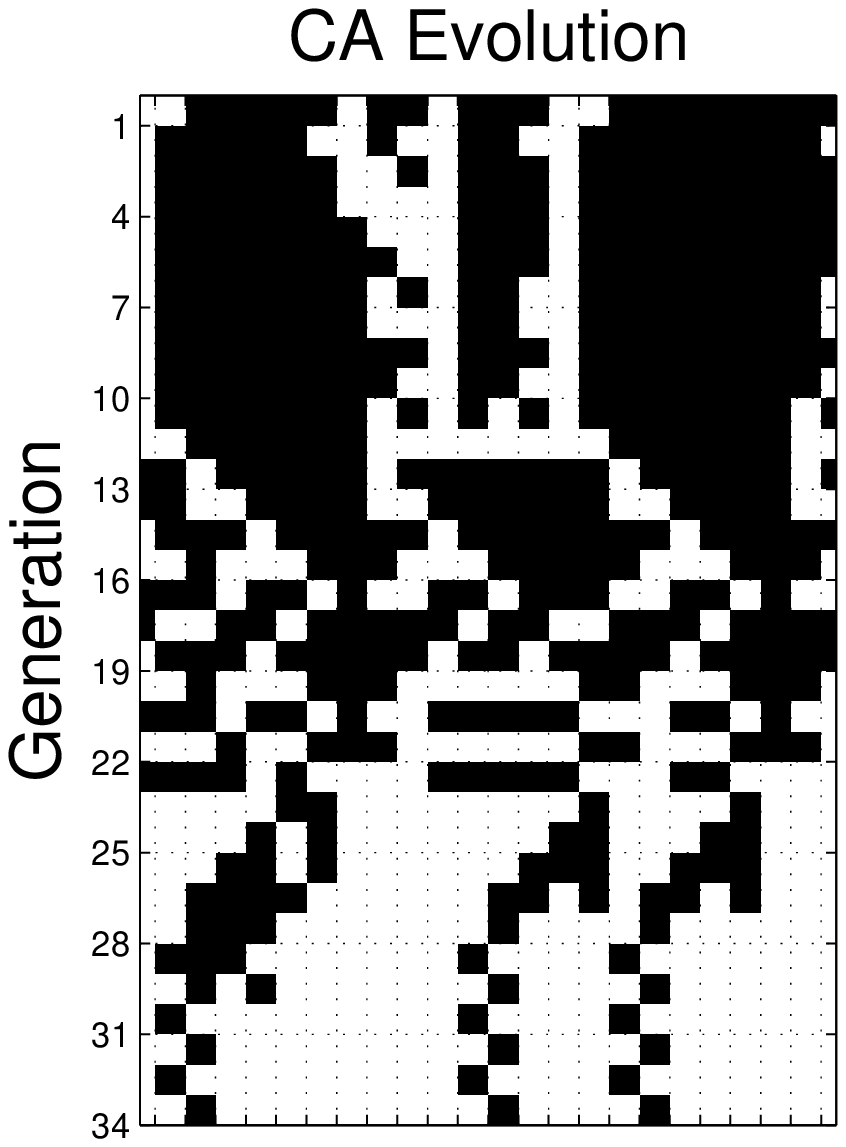}%
        \hfil
        \includegraphics[trim={0pt 30pt 70pt -10pt}, clip, height=0.25\textheight]{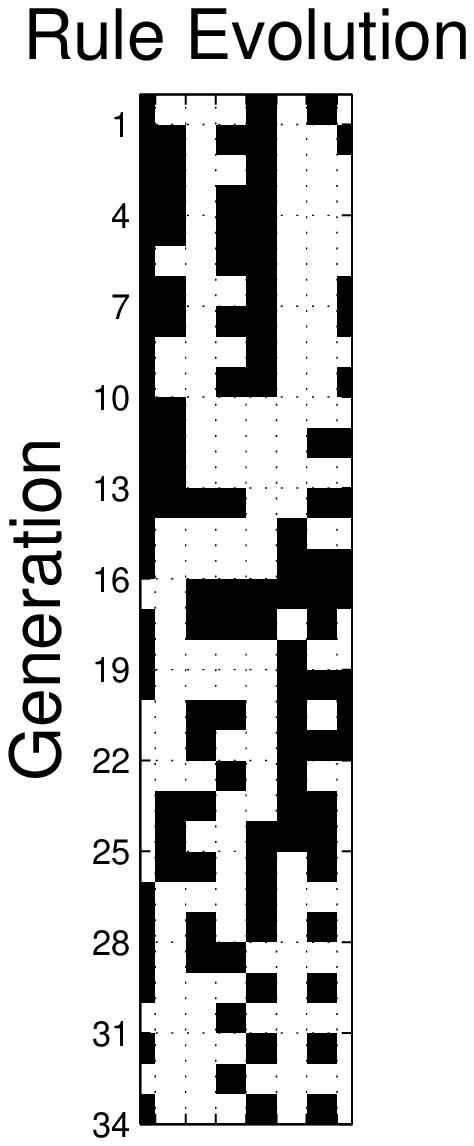}%
        }%
    }
    \subfigure[Execution leading to mutual fixed points]{%
        \label{fig:mutual_fixed}%
        \centerline{%
        \includegraphics[trim={0pt 30pt 0pt -10pt}, clip, height=0.25\textheight]{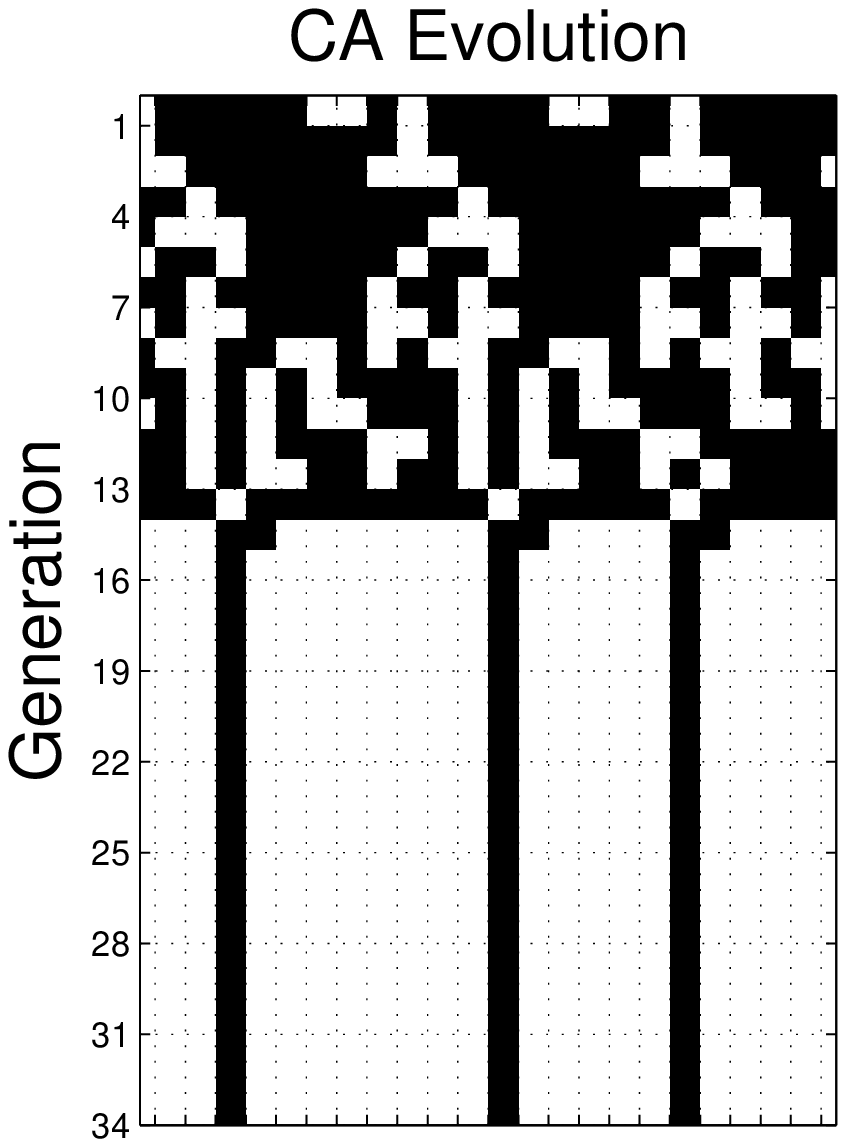}%
        \hfil
        \includegraphics[trim={0pt 30pt 70pt -10pt}, clip, height=0.25\textheight]{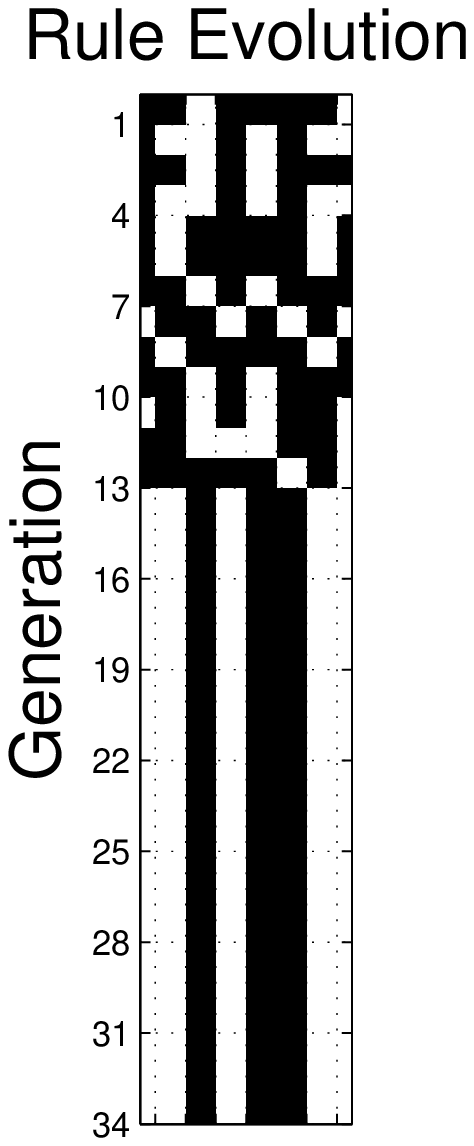}%
        }%
    }
    \subfigure[Microstate oscillations under invariant rule]{%
        \label{fig:invariance}%
        \centerline{%
        \includegraphics[trim={0pt 30pt 0pt -10pt}, clip, height=0.25\textheight]{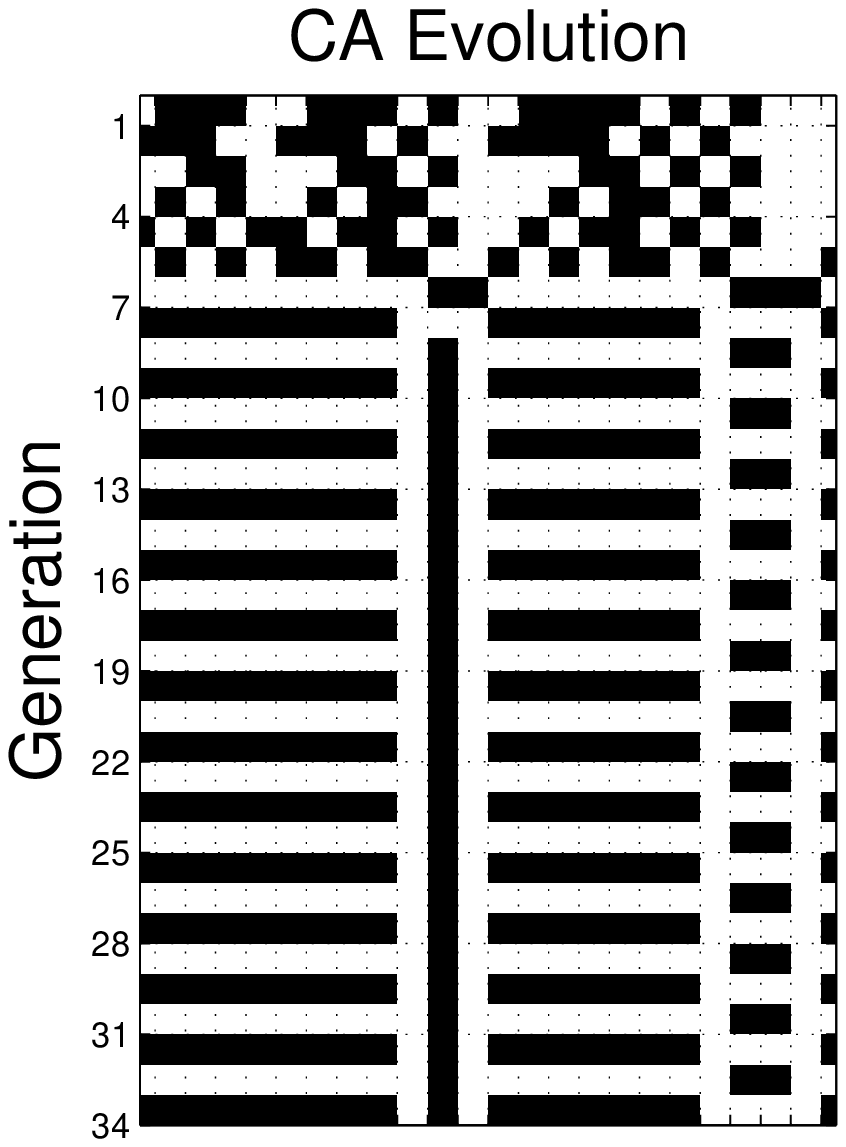}%
        \hfil
        \includegraphics[trim={0pt 30pt 70pt -10pt}, clip, height=0.25\textheight]{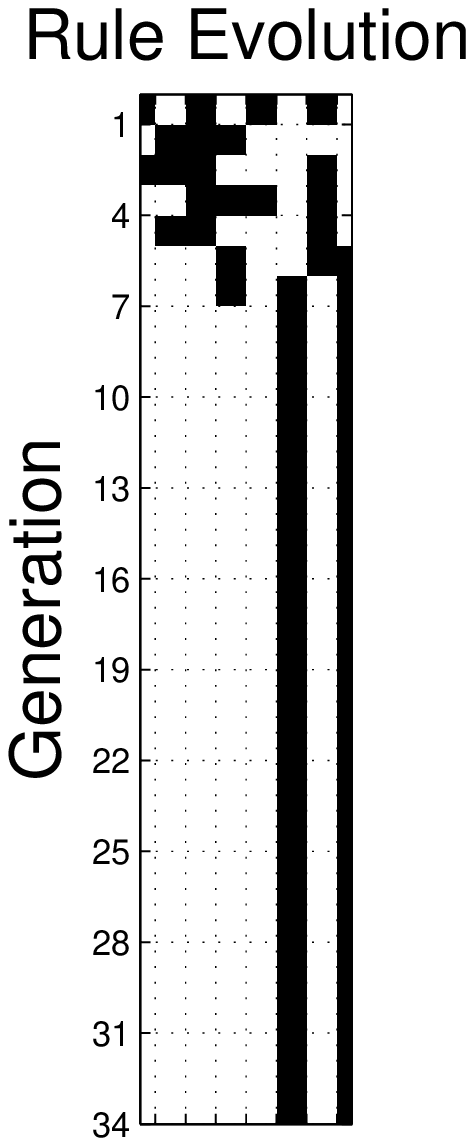}%
        }%
    }
    \caption
        [Example PICARD executions.]
        {Example PICARD executions. Cellular automata on the left are
        adjacent to their corresponding rule histories on the right.
        Executions in~\subref{fig:mutual_oscillation}
        and~\subref{fig:mutual_fixed} show CA and rule histories that
        have equilibria of the same type. In~\subref{fig:invariance}, an
        execution is shown where oscillations in microstate do not
        correspond to oscillations in rule.}
    \label{fig:example_picard_executions}
\end{figure}

In the following, we consider the mapping described above in
\autoref{fig:example_permuted_transition}. Because the rule induced by
each row can be viewed as a macrostate of the system, it is useful to
consider both the CA microstate dynamics and the rule macrostate
dynamics. In \autoref{fig:example_picard_executions}, executions of the
PICARD mapping are shown starting from three different initial
conditions. The executions of the CA are shown in the left column, and
the corresponding executions through the macrostate rule space is shown
in the right column.
\begin{figure*}\centering
    \includegraphics[trim={5pt 25pt 5pt 10pt}, width=\textwidth]{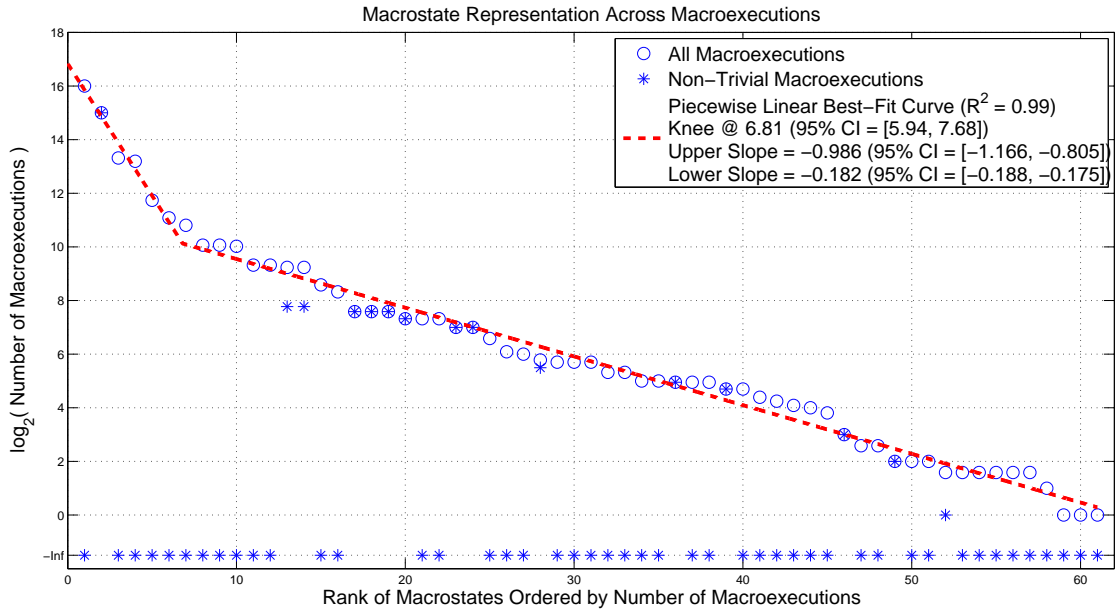}
    \caption
        [Macrostate representation over macroexecutions.]
        {Macrostate representation over macroexecutions. Each macrostate
         may be invariant across a range of disconnected
         macroexecutions. Every macroexecution terminates in a fixed
         point or a periodic cycle of microstates that each correspond
         to the same macrostate. Each terminating fixed point or cycle
         may be reached from a number of other transient microstates
         which themselves correspond to that same macrostate. A complete
         macroexecution includes the terminating cycle and all leading
         transient microstates which share the same macrostate. Using
         the PICARD mapping from
         \protect\autoref{fig:example_permuted_transition}, the entire
         microstate space has been explored and every complete
         macroexecution has been found. Displayed here as open circles
         are the $\log_2$\-/scaled counts of macroexecutions within each
         macrostate, omitting macrostates with no macroexecutions. The
         asterisks represent the $\log_2$-scaled count of
         macroexecutions after all fixed\-/point macroexecutions are
         removed. The horizontal axis shows the rank of the
         corresponding macrostate when ordered by the total number of
         complete macroexecutions within it. A broken piecewise\-/linear
         line of best fit is shown with an upper slope near $-1$.}
    \label{fig:macrostate_representation}
\end{figure*}
\begin{figure*}\centering
    \includegraphics[trim={5pt 25pt 5pt 10pt}, clip, width=\textwidth]{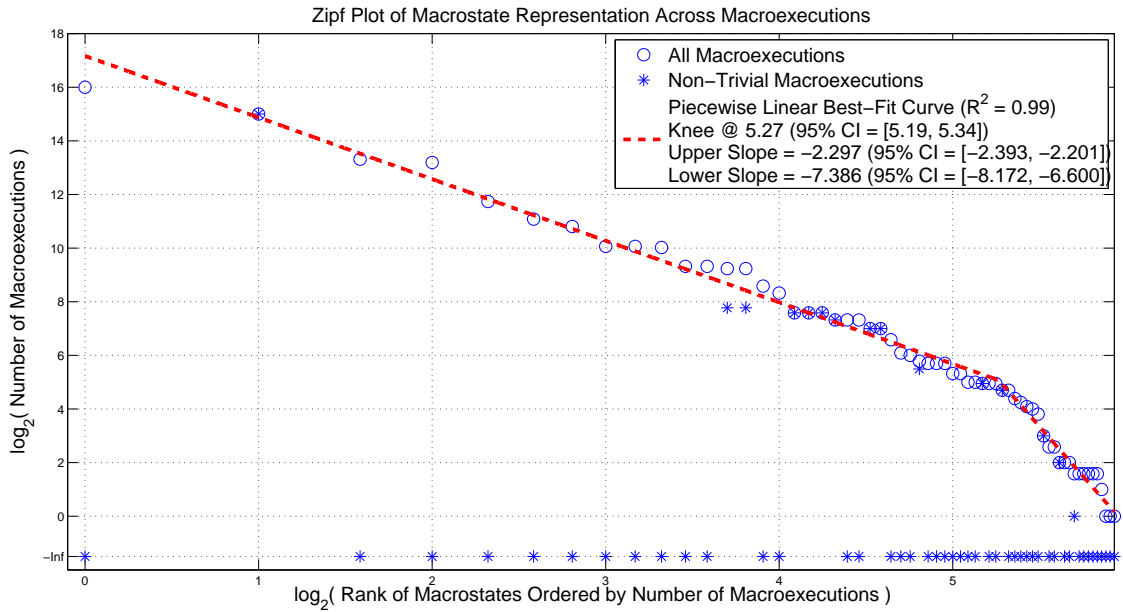}
    \caption
        [Zipf plot of macrostate distribution over macroexecutions.]
        {Zipf plot of macrostate distribution over macroexecutions.
         Displayed are the data from
         \protect\autoref{fig:macrostate_representation}; however,
         the horizontal axis is also $\log_2$ scaled. The broken
         piecewise\-/linear line of best fit shows that the distribution
         is approximately Zipfian.}
    \label{fig:zipf_plot}
\end{figure*}

In general, the lower resolution macrostate executions are each visually
similar to their higher resolution microstate. In
\autoref{fig:mutual_oscillation}, the microstate reaches a
steady\-/state oscillation. Given that the rule that governs the
microstate evolution is derived from the microstate itself, it is not
surprising that the macrostate also reaches a steady\-/state
oscillation. Similarly, in \autoref{fig:mutual_fixed}, the microstate
and macrostate both reach a fixed point. However, while the CA does not
reach its fixed point until generation~16, its macrostate becomes fixed
in generation~14. Thus, at generation~14, this PICARD degenerates into
an elementary CA under rule~44~(\ie, 0b00101100). If all generations
before~14 were removed from the history of the CA, it would be
indistinguishable from an elementary CA. Moreover, as shown in
\autoref{fig:invariance}, this phenomenon is not restricted to only
fixed points. By generation~8, the rule trajectory becomes fixed on
rule~5~(\ie, 0b00000101); however, the CA trajectory reaches a
steady\-/state oscillation where some cells are fixed and others cycle.
Again, the tail of this CA execution is indistinguishable from one that
would be generated under elementary CA rule~5. Consequently, this single
PICARD mapping is able to represent executions from multiple elementary
CA rules simultaneously while also potentially producing novel
executions.

When a PICARD mapping generates an execution that is in a region of
macrostate invariance, we call that execution a \emph{macroexecution.}
Once a macroexecution has been reached, state feedback is not required
to determine the future trajectory. Thus, macroexecutions move through
regions of \emph{locally elementary} CA space. For the mapping explored
in this section, \autoref{fig:macrostate_representation} shows the
relative ``size'' of each of these locally elementary regions.
In particular, for each of the 256 elementary CA rules, all of the
disconnected macroexecutions for which the rule is invariant were
counted. The rules with non\-/zero counts were ranked and plotted as the
open circles in the semilog frequency distribution in
\autoref{fig:macrostate_representation}. The asterisks on the plot
represent the number of oscillatory macroexecutions~-- that is, the
asterisks represent the count when the fixed\-/point macroexecutions are
removed. Thus, the highest ranking elementary CA rule includes all
fixed\-/point macroexecutions, but the second ranking elementary CA rule
includes no fixed\-/point macroexecutions.

As shown by the piecewise linear broken line in
\autoref{fig:macrostate_representation}, the highest ranking elementary
CA rules have counts which peak at $2^{16}$ and are halved with each
increase in rank over a significant range of rules. However, this trend
is not long lasting and may simply be an artifact of the particular
PICARD mapping we have chosen for this case study. In
\autoref{fig:zipf_plot}, the same distribution is shown on a log--log
plot and shows evidence that the greater trend is
Zipfian~\citep{Zipf49}.
Words in a language tend to be Zipfian distributed in their use~-- the
relative frequency of words is inversely proportional to their rank.
Following an analogy with language, elementary CA rules act like words
which are used in PICARD macroexecutions. Using a functional\-/dynamics
framework as a model for the evolution of natural language,
\citet{KK00b} show how self\-/referencing function dynamics can act like
a filter that produces a corpus of words that are each fixed points of
the dynamical model. The invariant macrostate CA rules that we describe
are qualitatively similar to this idea, and their representation is
consistent with measured distributions of words in actual natural
language.

\begin{figure}\centering
    \centerline{%
    \includegraphics[trim={0pt 30pt 0pt -10pt}, clip, height=0.25\textheight]{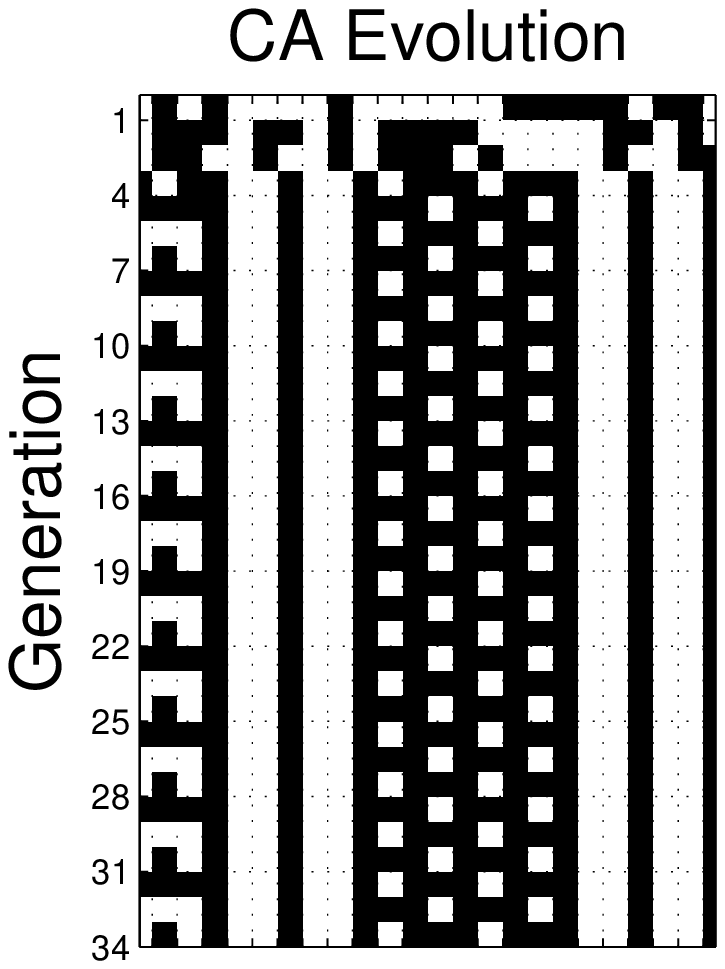}%
    \hfil
    \includegraphics[trim={0pt 30pt 50pt -10pt}, clip, height=0.25\textheight]{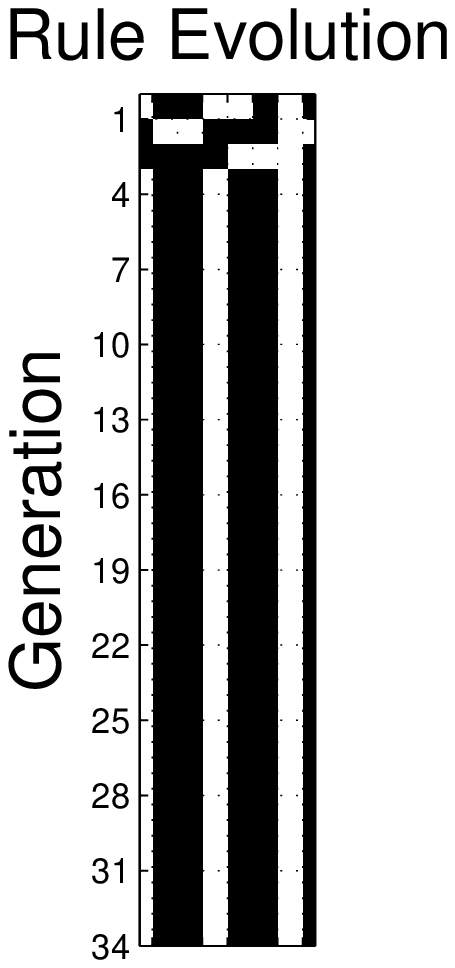}%
    }%
    \caption
        [Period\-/6 macroexecution with short transient.]
        {Period\-/6 macroexecution with short transient.
         A steady\-/state rule-109~(0x6D) macroexecution is shown that
         has a period of six generations preceded by its longest
         possible transient. Rule 109 has 31 macroexecutions, which are
         all non\-/trivial.}
    \label{fig:period_six_oscillation}
\end{figure}

The data from \autoref{fig:macrostate_representation} have been
reproduced in \autoref{tab:high_ranking_macrostates} for the sixteen
highest\-/ranking macrostates.
\begin{table}\centering
    \begin{tabular}{rcc}
        \toprule
        \multicolumn{1}{c}{Rule} & Macroexecutions & \shortstack{Non\-/trivial\\Macroexecutions}\\
        \midrule
        204 (0xCC) & 65536 & 0\\
        51 (0x33) & 32768 & 32768\\
        205 (0xCD) & 10196 & 0\\
        76 (0x4C) & 9360 & 0\\
        236 (0xEC) & 3420 & 0\\
        200 (0xC8) & 2166 & 0\\
        4 (0x04) & 1792 & 0\\
        12 (0x0C) & 1072 & 0\\
        132 (0x84) & 1072 & 0\\
        68 (0x44) & 1040 & 0\\
        140 (0x8C) & 640 & 0\\
        196 (0xC4) & 640 & 0\\
        108 (0x6C) & 603 & 219\\
        201 (0xC9) & 603 & 219\\
        232 (0xE8) & 384 & 0\\
        77 (0x4D) & 320 & 0\\
        \bottomrule
    \end{tabular}
    \caption
        [Sixteen highest ranking macrostates.]
        {Sixteen highest ranking macrostates.}
    \label{tab:high_ranking_macrostates}.
\end{table}
The two highest\-/ranking macrostates, rule~204 and rule~51, each have a
number of macroexecutions that is a power of two, $2^{16}$ and $2^{15}$.
In the case of rule~204, its macroexecutions exhaust all of the $2^{16}$
microstates that map to it. Thus, rule~204 has no transients~-- it is a
region filled entirely with fixed points. Likewise, because there are no
fixed points in rule~51 and yet $2^{15}$ macroexecutions, each
macroexecution must be a period\-/2 cycle. So the rule-51 region also
has no transient microstates, but it is filled entirely with these
cycles. Moving farther down the ranks, more microstates become available
for forming macroexecutions with longer\-/period oscillations within
them. For example, \autoref{fig:period_six_oscillation} shows a
period\-/6 oscillation embedded within a rule-109 macroexecution.
Because each of the $2^8$ macrostates represents a finite number of
microstates ($2^{16}$), the upper bound on the number of distinct
oscillations which are macrostate invariant must decrease with the
average length of those oscillations. Likewise, rule~109 has only 31
macroexecutions, which are all non\-/trivial and thus all end on cycles
of period 2 or greater. Although this mapping favors short\-/period
oscillations, other mappings may allow for longer cycles. For example,
if the aggregate sum mapping in \autoref{fig:example_mapping_sum} is
composed with a function that maps a unity sum to rule~4, then PICARD
will shift each sum-1 row to the right by one position on each
iteration. These shifts will not alter the sum of the row, and thus this
shifting will continue indefinitely. Consequently, in this hypothetical
case, the rule-4 region will contain macroexecutions with periods as
long as the length of each microstate row. However, by the discussion
above, rule~4 is unlikely to contain a high number of macroexecutions
due to the number of microstates consumed by these long\-/period
oscillations.

\begin{figure}\centering
    \centerline{%
    \includegraphics[trim={0pt 30pt 0pt -10pt}, clip, height=0.25\textheight]{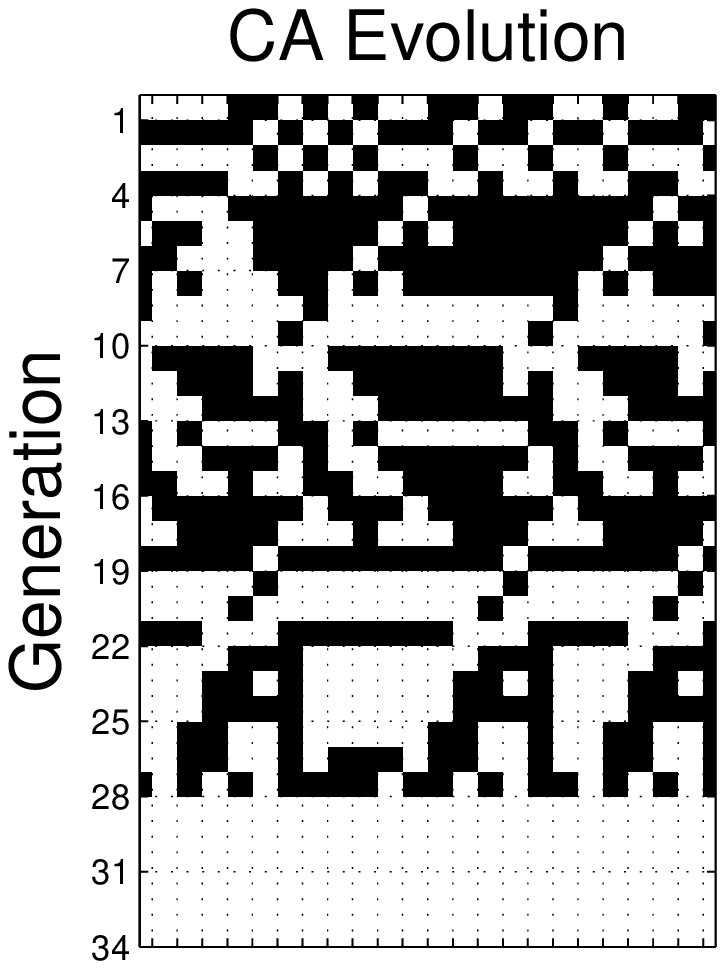}%
    \hfil
    \includegraphics[trim={0pt 30pt 50pt -10pt}, clip, height=0.25\textheight]{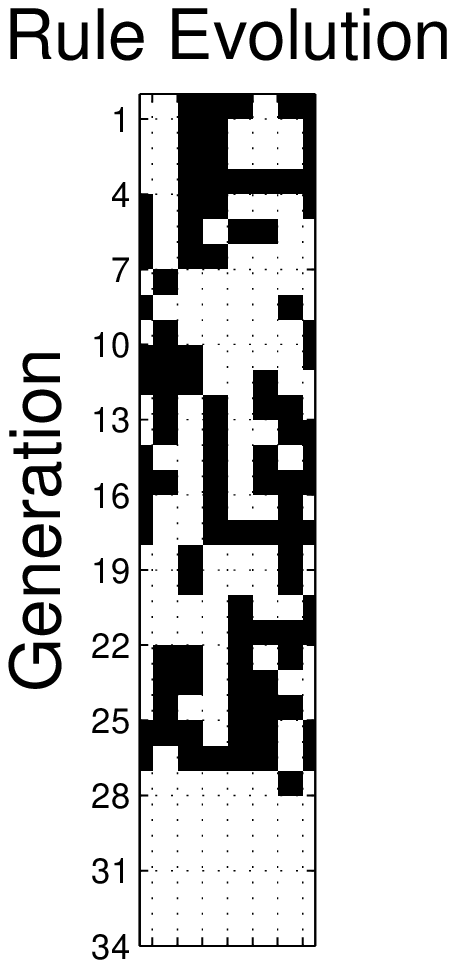}%
    }%
    \caption
        [Fixed point with long structured transient.]
        {Fixed point with long structured transient.
         A rule-0 fixed point is
         shown that is preceded by a very long transient that appears to
         have significant visual structure.}
    \label{fig:long_ugly_transient}%
\end{figure}

As shown above in \autoref{fig:mutual_oscillation}, PICARD oscillations
need not be contained within a macroexecution~-- microstates can
oscillate across rule boundaries. Additionally, the transient components
of PICARD executions can have rich structure. In
\autoref{fig:long_ugly_transient}, there is a long sequence of transient
execution leading up to an eventual rule-0 fixed point.
The CA appears to have significant structure visually, but the rule
evolution is less predictable and shows clear path dependence. Thus, the
feedback in PICARD both constrains its executions and generates novel
patterns that may be fodder for further analysis.

\section{Summary and Future Work}

By feeding a conventional one\-/dimensional elementary cellular
automaton's own state back into the iteration rule that generates its
next state, we have developed a new modeling framework for investigating
information control in natural systems. Although open\-/loop cellular
automata that evolve under strictly static rules have been useful tools
for studying evolution artificially, these classic frameworks are not
sufficiently rich to model the dependence of fitness on current state.
Moreover, by adapting the methods of evolving cellular automata so that
PICARD mappings can be the targets of artificial selection, there is a
potentially richer set of behaviors that can be selected for.

We have shown how state feedback can generate self\-/reinforcing regions
of behavior. Thus, this framework provides a model of how functional
diversity can be embedded within a single automaton. This functional
diversity is similar to the diverse differentiation possibilities for
cells in living organisms. Moreover, as evidenced by the distribution of
macrostates over macroexecutions, this framework apparently shares
similarities with the processes responsible for the generation of
natural language.

A possible criticism of PICARD is that it introduces non\-/local effects
to cellular automata. Traditional CA are built from the assumptions that
cells update based entirely on local information and information about
neighbors. By feeding the state of an entire row back into the CA rule,
this locality assumption is broken. A complete reductionist model of the
dynamics of life would necessarily have to incorporate dynamics of both
the focal entity and the environment around it~-- despite any
significant differences in time scales. By feeding the state of the
focal entity back into the rules that govern how the entity evolves, we
avoid these complications and still allow for entities to alter
\emph{and be altered by} their environment. Thus, just as
coarse\-/grained descriptions have utility in modeling physical
phenomena, non\-/local effects in PICARD mappings can be viewed as
useful coarse\-/grainings of otherwise less tractable models.

As PICARD Implements CA Rules Differently, there are myriad directions
for future work~-- including directions which parallel investigations
from conventional cellular automata and directions which are specific to
PICARD self\-/reference. At a minimum, the distribution of macrostates
over macroexecutions needs to be investigated for a wide variety of
PICARD mappings to determine whether the Zipfian distribution is
widespread. Although it is attractive to view the PICARD automata in
this paper as a mosaic of different elementary CA that each govern small
patches of local elementarity, it is unlikely that such regions exist
that contain executions with the relatively open\-/ended complexity of
open\-/loop elementary CA patterns. However, PICARD mappings have the
potential of generating new patterns and new functionality that may be
out of reach of conventional CA's. Alternatively, PICARD mappings may
have the ability to increase the robustness of elementary CA function by
reducing the sensitivity to minute but functionally inconsequential
changes in initial condition. Extending PICARD mappings to be able to
map from prior microstates may allow for incorporating functionality
normally associated with higher\-/dimensional cellular automata. For
example, \citeauthor{Langton84}'s loops~\citep{Langton84} are able to
self replicate because growing patterns in space can turn and interact
with earlier patterns. A PICARD mapping can similarly connect iterations
across time and draw a connection between two\-/dimensional self
replication and one\-/dimensional pattern generation. In general, an
important future direction is to connect one\-/dimensional PICARD
insights to higher dimensional cellular automata.

\section{Acknowledgements}

Kunihiko Kaneko and Larissa Albantakis provided useful feedback on
earlier versions of this work. Interaction with them was possible due to
a workshop on Information, Complexity, and Life organized by the BEYOND
Center for Fundamental Concepts in Science at Arizona State University.
Andrea Richa also provided helpful comments regarding the presentation
of this material. This project/publication was made possible through
support of a grant from Templeton World Charity Foundation. The
opinions expressed in this publication are those of the author(s) and do
not necessarily reflect the views of Templeton World Charity Foundation.


\footnotesize
\bibliography{self_ref_ca}

\end{document}